\documentclass[12pt,english]{article}
\usepackage[T1]{fontenc}
\usepackage[latin9]{inputenc}
\usepackage{natbib}
\usepackage{csquotes}
\usepackage{setspace}
\usepackage{babel}
\usepackage[text={16.5cm,23cm},centering]{geometry}
\expandafter\def\expandafter\quote\expandafter{\quote\small}

\makeatletter

\@ifundefined{date}{}{\date{}}
\makeatother

\begin{document}

\title{\textbf{A fuzzy process of individuation}}

\author{Juliano C. S. Neves%
\thanks{nevesjcs@if.usp.br%
}}

\maketitle

\begin{center}
{\it{Centro de Ciências Naturais e Humanas, Universidade Federal do ABC,\\ Avenida dos Estados 5001, Santo André, 09210-580 São Paulo, Brazil}}
\end{center}

\vspace{0.5cm}

\begin{abstract}
It is shown that an aspect of the process of individuation may be thought of as a fuzzy set. The process of individuation has been interpreted as a two-valued problem in the history of philosophy. In this work, I intend to show that such a process in its psychosocial aspect is better understood in terms of a fuzzy set, characterized by a continuum membership function. According to this perspective, species and their members present different degrees of individuation. Such degrees are measured from the membership function of the psychosocial process of individuation. Thus, a social analysis is suggested by using this approach in human societies. 
\end{abstract}


\section{Introduction}
Fuzzy sets were a revolution in the 20th century. Lotfi Zadeh's work (\citealt{Zadeh}) has been applied in several areas in science and technology.\footnote{An introduction to fuzzy sets and its applications is found in \cite{Zimmermann1,Zimmermann2}.} The main feature of fuzziness is the ability of characterizing process, systems, and events with the aid of a continuum function, the membership function. Such a function describes the continuum degree of membership of the studied system to a specific fuzzy set. That is, within the fuzzy set approach, responses are not given only in terms of \enquote{0} or \enquote{1}, \enquote{yes} or \enquote{no}, and \enquote{true} or \enquote{false}. There are degrees of membership. This important feature of the fuzzy sets will be used in order to describe mathematically the process of individuation, which provides the notion of individual in philosophy. Then, mathematically, the question about individuation will not be raised in terms of the dichotomy \enquote{non-individual} or \enquote{individual} described by an ordinary set (also called crisp set) or a binary reasoning. I will describe the process of individuation---at least its psychosocial aspect---in terms of the degree of individuation, i.e., a process of individuation will be defined as a fuzzy set. It is worth noting that fuzzy sets have been applied to social sciences and psychology as well. For example, \cite{Leung} described  the degree of liking between individuals by using a continuum membership function, and \cite{Smithson} studied fuzziness in behavioral sciences.      

According to several philosophers, individuation has been thought of as a binary variable in the history of philosophy. For those who thought of individuation as a process, like the French philosopher Gilbert \cite{Simondon}, individuation, or an individual, is considered as result of a process, and non-individuation is the lack of such a process. However, as indicated by Simondon in the 20th century, among individuals, especially living beings, it is not hard to see differences of degrees of individuation. In living beings,  I suggest that degrees of individuation  are given by a fuzzy set. Then, the process of individuation and, as we will see, even the sense of belonging to a species or a larger group and structure may be thought of in terms of  fuzzy sets in a psychosocial perspective. This perspective of  the process of individuation---the psychosocial one---will be focused on this article. I will describe it mathematically by means of the membership function of fuzzy sets. Thus, the process of individuation and the sense of belonging will not be 
 described by means of a binary variable. 

For important scholars in the history of philosophy (\citealt{Various}), individuation is supposedly generated by the principle of individuation. But, following Simondon and rejecting a metaphysical principle, one may indicate individuation as a process and describe it physically and mathematically in terms of physical quantities and by using a continuum membership function. The psychosocial aspect of the individuated being 
will be  conceived of as fuzzy sets endowed with continuum membership functions that provide the grade of individuation
 and the grade of  belonging of a specific individual to a larger structure. By using this approach
  in order to describe the process of individuation in its psychosocial aspect, it is  shown that the process of
   individuation, as a non-binary  process, can be applied to characterize human societies.    

The structure of this article is as follows: Sec. 2 presents the definitions of both the principle of individuation and process of individuation, focusing on the psychosocial aspect; in Sec. 3 the fuzzy process of individuation is defined with a suggested application in human societies in Sec. 4. The final remarks are presented in Sec. 5.

\section{The process of individuation}
In philosophy, the principle of individuation (\textit{principium individuationis}) has been a very useful concept. According to dictionaries of philosophy, the principle of individuation is \enquote{what makes something individual as opposed to universal} \citep[p. 737]{Various}. In this sense, a specific man is different from the concept, or universal man, because of the principle of individuation. Therefore, in this perspective, our world is described as multiplicity of entities because the principle of individuation is acting. 

In Arthur Schopenhauer's philosophy, for example, the principle of individuation promotes the world as representation. As the being-in-itself or the thing-in-itself is the \textit{will} (beyond the principle of individuation, that in his philosophy is equivalent to space and time\footnote{\enquote{I shall call time and space the \textit{principium individuationis}, an expression borrowed from the old scholasticism (...)}, \cite[Second Book, \S 23]{Schopenhauer}.}), Schopenhauer says that the principle of individuation generates individuals from the will (unity), which is the origin of the world, \enquote{the innermost essence, the kernel, of every particular thing and also of the whole.}\footnote{\citet[Second book, \S 21]{Schopenhauer}.} Therefore, unity, or the thing-in-itself, presents itself as a myriad of objects, i.e., the world, our sensible or physical world, is the \textit{will} by means of the principle of individuation.

On the other hand, for Friedrich Nietzsche, at least in the very first period of his work, the principle of individuation is identified to a drive (\textit{Trieb}), which received the name of the Greek God Apollo. But the origin of the world, such as in Schopenhauer's work, is attributed to unity, which in Nietzsche's initial philosophy received the name of primordial unity (\textit{Ur-Eine}).\footnote{\citet[I]{Tragedy}. The mature Nietzsche rejected the primordial unity because his philosophy of wills to power is plural.} Above all, the principle of individuation for Nietzsche and Schopenhauer was the attempt of describing the multiplicity in terms of a unique origin or a unique cause. Individuation and its supposed cause, the principle of individuation, were ingredients to justify the sensible world from a single metaphysical origin.

In this work, by ignoring metaphysical speculations of a supposed principle of individuation, I conceive of individuation among and within species as a natural process. It is not hard to see that individuation is promoted in different degrees, according to species. Each member, or individual, recognizes in some degree the difference between the \enquote{self} and the world, the \enquote{self} and other objects and members of species or community. There exists an enormous difference in the process of individuation if one considers different species. For example, individuation is more stressed in men than in ants. Even in human societies, the process of individuation presents different degrees. The sense of participation or the sense of belonging to a larger entity may decrease the degree of individuation in some cases. According to Nietzsche, for example, there were elements in the Greek tragedy that ensured the belief in some kind of participation in something higher, there was a \enquote{sense of belonging to a higher community.}\footnote{Ibidem.} Members of the audience recognized a union between people and nature. This return to nature was promoted by another drive in Nietzschean philosophy---the Greek God Dionysus. That is, individuals, or citizens, during the Greek tragedy had their degrees of individuation decreased.

Both the process of individuation and the sense of belonging are interpreted as natural drives in this article.\footnote{Without Schopenhauer's influence, the mature Nietzsche constructed his own work. In that period, the Apolline and Dionysian drives are not abandoned. From that period, they carry a totally natural sense: they are manifestations of wills to power.} I will focus on the psychosocial feature of the process of individuation (and, consequently, on the sense of belonging). And we will find the notion of psychosocial process of individuation in Simondon's work.  

In the 20th century, considering the importance of looking at \textit{degrees of individuation} as well,\footnote{\enquote{I intend therefore to study the \textit{forms, modes and degrees of individuation} in order to situate accurately the individual in the wider being according to the three levels of the physical, the vital and the psychosocial,} \citet[p. 311]{Simondon}. As we can see, there are degrees, which will be interpreted as continuum in the fuzzy approach, and three levels of individuation as well, which will be discussed hereafter.} Simondon (\citealt{Simondon,Simondon2}) wrote about the process of individuation, or ontogenesis, as something better understood in terms of a non-classical logic, instead of the binary one: \enquote{one sees that classical logic can not be used to understand individuation (...)}.\footnote{Idem, p. 312.  A possible non-classical logic is the fuzzy logic, which comes from the fuzzy sets.} Simondon presented in his philosophy of individuation an influential interpretation of the process of individuation. For him, the process of individuation is better described by not only a non-classical logic, but using modern physical concepts as well, like energy and (meta)-stability. According to Simondon, modern physics offers adequate tools in order to interpret individuals as process, which are systems with nonvanishing potential energies. In particular, an individual is the process of individuation acting, is not a fixed result of a process. For living beings (and Simondon's perspective describes all individuals, living beings or not), the process of individuation is something that occurs between the individual and the environment (\textit{milieu}):\footnote{Idem, p. 300.}
\begin{quote}
The process of individuation must be considered primordial, for it is this process that at once brings the individual into being and determines all the distinguishing characteristics of its development, organization and modalities. Thus, the individual is to be understood as having a relative reality, occupying only a certain phase of the whole being in question -- a phase that therefore carries the implication of a preceding preindividual state, and that, even after individuation, does not exist in isolation, since individuation does not exhaust in the single act of its appearance all the potentials embedded in the preindividual state. Individuation, moreover, not only brings the individual to light but also the individual-milieu dyad.
\end{quote}     
The individuated being, or individual, according to Simondon, is a metastable state---a dynamical process that solves problems, acts, and simultaneously individuates itself. The individual is a state that appears from a preindividual state. From this perspective, the terminology used by Simondon---the concepts of process and state---is very adequate. 
Because the individuated being is not a static being, a final state, something eternal and firm. In my reading, Simondon and his philosophy of individuation bring concepts and similar interpretations to Heraclitus and Nietzsche. Such as Heraclitus, one can see the importance of becoming, or process. Such as Nietzsche, one can find the question about \enquote{stability} of individuals. For Nietzsche, individuated beings are transitory configurations of wills to power.\footnote{See \citet{Muller}.} In particular, man is not an \textit{aeterna veritas} (eternal truth):\footnote{\citet[p. 12]{Human}.}
\begin{quote}
All philosophers have the common failing of starting out from man as he is now and thinking they can reach their goal through an analysis of him. They involuntarily think of \enquote{man} as an \textit{aeterna veritas}, as something that remains constant in the midst of all flux, as a sure measure of things.
\end{quote}

Simondon considered levels of individuation as well. The process of individuation is origin of both physical objects and living beings. For living beings, specifically, individuation is more sophisticated: \enquote{The living being resolves its problems not only by adapting itself which is to say, by modifying its relationship to its milieu (something a machine is equally able to do) -- but by modifying itself through the invention of new internal structures (...)}.\footnote{\citet[p. 305]{Simondon}.} For Simondon, living beings participate of the process of individuation in \textit{three levels}: physical, vital and psychosocial level. The physical one is considered because living beings are \enquote{made up} of particles (or quantum fields), according to the standard model of particles. The vital level represents the biological individuation. And the last one---the psychosocial individuation---is the union of both the psychological and social individuation. In Simondon philosophy, there is no strict separation between social and psychological processes. In this sense, the fuzzy process of individuation proposed in the next section captures the degree of psychological and social individuation. 
 The fuzzy approach also provides the increased sense of belonging  indicated by Nietzsche in the Greek
  tragedy. Accordingly, belonging and individuation (or individuality)  can be described as fuzzy sets, providing
   the notions of degree of belonging and degree of individuality.   
 
Following Simondon and a natural approach for the process of individuation to the detriment of the metaphysical one, \cite{Weinbaum} proposed recently a new form to define and conceive intelligence by means of the process of individuation. According to the authors, intelligent agents emerge from a complex context and become intelligent from a process of self-organization and formation, where \enquote{individuation is a resolution of a problematic situation} \citep[p. 10]{Weinbaum}. As we can see, the process of individuation is totally naturalized for those authors as well.        

In the next section, I will indicate a different approach to the process of individuation. Rather than a binary process, where 0 means non-individual, and 1 means a full individual, I will adopt a fuzzy perspective.\footnote{It is worth mentioning that for Schopenhauer there were degrees of manifestation of the will. Men are higher degrees of manifestation, or higher degrees of objectivity, than ants. Then, in my interpretation, one can see indications for a fuzzy approach in Schopenhauer philosophy as well (see \citealt[Second book, \S 21 and \S 24]{Schopenhauer}). A process of individuation with degrees of individuation was indicated in \cite{Neves} (and as we saw in \cite{Simondon}), but without a formulation in terms of fuzzy sets.} Then, the process of individuation in its psychosocial aspect will be thought of as a fuzzy set, characterized by a membership function, which assumes a continuum interval of values between 0 and 1.

\section{Describing a fuzzy process of individuation}
To think about a fuzzy process of individuation means to realize various degrees of individuation among and within species. The process of individuation in its psychosocial aspect may be thought of as a subset $P$ of $X$ ($P \subseteq X$), where $X$ is the set of all individuals. Each individual of $X$, a member of a species, group or society, is indicated by $x$. One assumes that $P$ is a fuzzy set characterized by a membership function $f_{P}(x)$. Then the membership function for each individual in $X$ is described by a continuum interval $[0,1]$. For each individual, the membership function measures the degree of psychosocial individuation: the minimum value is 0, and the maximum is 1. Therefore, the fuzzy psychosocial process of individuation $P$ is defined as set of ordered pairs $(x,f_P(x))$ such that
\begin{equation}
P = \lbrace (x,f_P(x)) : x \in X, f_{P}(x) : X \rightarrow [0,1] \rbrace.
\label{function}
\end{equation}
If I had used an ordinary (or crisp) set in order to describe $P$, the membership function would assume only 0 or 1 ($ f_{P}(x) : X \rightarrow \lbrace 0,1 \rbrace$). That is, in ordinary sets theory, an individual $x$ would be or would not be included in $P$. But from the definition of the psychosocial process of individuation, given by Eq. (\ref{function}), there will be degrees of membership in $P$, not only two possible answers.
 
The extreme values $0$ and $1$ for the fuzzy process of individuation have interesting meanings. The value $0$ means absence of individuality, and if the membership function is $f_P(x)=0$ for all $x's$, we will have the empty set $\emptyset$ of the psychosocial process of individuation:
\begin{equation}
\emptyset =  \lbrace ( x, f_P(x)=0),\forall x \in X  \rbrace.
\end{equation}
Therefore, there is no social individual, and the psychosocial process of individuation is not acting. On the other hand, the value $1$ for the membership function stands for a full or complete individual---the extremal case of the
 psychosocial individuation. If the membership function is $f_P(x)=1$ for all $x's$, we will have the universal fuzzy set $U$ of the psychosocial process of individuation written as
\begin{equation}
U= \lbrace ( x, f_P(x)=1),\forall x \in X  \rbrace.
\end{equation}
The extreme fuzzy sets $\emptyset$ and $U$ have the following relation with the fuzzy set $P$:
\begin{equation}
\emptyset \leq P \leq U.
\end{equation}

Thus, the maximum and minimum values of $f_P(x)$ mean: \textit{the closer to 0, the smaller the
 psychosocial process of individuation, and the closer to 1, the greater the psychosocial process of individuation}.  
 As I said, the sense of belonging is also interpreted within this approach.
  Belonging $B$ can be conceived of as a fuzzy set, and given an individual $x$, he/she will have a degree of  
   membership in $B$, which is defined as set of ordered pairs $(x,f_B(x))$, i.e.,  
   \begin{equation}
B = \lbrace (x,f_B(x)) : x \in X, f_{B}(x) : X \rightarrow [0,1] \rbrace.
\label{B}
\end{equation}
 An individual presents degrees of membership in $P$ and $B$ at the same time, has a certain degree of individuation
  and a  certain degree of belonging to a group. In general, $P$ and $B$ will not be conceived of as correlated
   sets \textit{a priori}.
   That is, high degree of individuation does not necessarily mean low degree of belonging and vice versa.
    However, in some cases,
    for example based on Nietzsche's interpretation of Greek tragedy, the concepts of individuation and belonging may be
     negatively correlated to each other. 
    As Nietzsche pointed out, Greek tragedy increased the sense of belonging and decreased
     the sense of individuality or individuation of the audience. But, as I said,
        this is not the general case, then the process of individuation and the sense of belonging are not necessarily
         assumed to be negatively correlated or complements of each other.

Among and within species, the value of $f_P(x)$ is not constant. That is, within a fixed species $S$, the membership function assumes
\begin{equation}
\min_{x \in S} f_{P} (x)< f_{P}(x) < \max_{x \in S} f_{P} (x).
\end{equation} 
Then, there are individuals with a larger degree of  individuation than others. Even for a specific individual, $f_P(x)$ is not constant along his/her entire life. Comparing different species, the difference in the process of individuation is even more notable. For example, I suggest that the values $\min f_{P}(x)$ and $\max f_{P}(x)$ in humans are larger than in ants. Therefore, when we compare species, groups or societies, it is more appropriate to assume $f_P(x)$ as a random variable and obtain the usual mean value of the process of individuation, $\bar{f}_P(S)$, for each species, group or society:
\begin{equation}
\bar{f}_P(S)=\frac{1}{N} \sum_{i=1}^{N}f_P(x_i),
\end{equation}
where $N$ stands for the number of individuals of a specific society, group, and species ($x_1,x_2,...,x_N$). That is, for humans $(H)$ and ants $(A)$, for example, one has the following relation for the mean value of the process of individuation:
\begin{equation}
\bar{ f}_{P}(H) >\bar{f}_{P}(A), 
\end{equation}
which reflects the uniformity of the social life for a large number of individuals in a society of ants.  

As pointed out earlier, there are intervals of individuation for each species or group. Then, the values of $f_{P}(x)$, or the values of individuation, may assume different intervals according to specific human societies. Besides, one can assume that the human interval, due to the cultural variations, is larger than the corresponding interval for any species. Comparing both the human and ant species once again, one has
\begin{equation}
\Delta f_P(H) > \Delta f_P(A),
\end{equation}
where $\Delta f_P(x)$ stands for $\max f_{P}(x) - \min f_{P}(x)$.

\section{A suggested social analysis}
Following Nietzsche, who characterized societies from human types,\footnote{In \citet[p. 84, fragment 2 (127) of 1885]{Fragments}, for example, it is found \enquote{the nihilist consequences of the political and economic way of thinking}. For the German philosopher, nihilistic individuals generate nihilistic societies and vice versa. Using Simondon's framework, both systems, individual and society, are not stable, they interact with themselves. For Simondon, there are processes of individuation for societies as well: \enquote{Individuation in its collective aspect makes a group individual (...)} (\citealt[p. 307]{Simondon}). Both processes, individual and social, are dynamical, generating mutual influences.} I suggest a statistical application of the fuzzy process of individuation in order to describe societies, focusing on human societies. From the above discussion, I propose that the psychosocial process of individuation, for example, is more stressed in capitalist societies than in socialist ones, i.e., the mean value of $f_{P}(x)$ among citizens is larger in the capitalist system than in the socialist system. Moreover, the sense of belonging to a higher structure---the modern state---is more emphasized by the people in socialist societies.

 A low mean value for $f_{P}(x)$ indicates low degrees of individuation among citizens. For Nietzsche, controversially, socialism \enquote{aspires to the annihilation of the individual}\footnote{\citet[p. 173]{Human}.} because the socialist state is more dominant in the social life. That is, socialism \enquote{desires an abundance of state power}.\footnote{Ibidem.} In another text, the German philosopher relates both the state development and the individual: \enquote{The state is a prudent institution for the protection of individuals against one another: if it is completed and perfected too far it will in the end enfeeble the individual and, indeed, dissolve him (...)}.\footnote{Idem, p. 113.} The strong presence of the socialist state imposes a lower degree of individuation in socialist societies. 

On the other hand, in capitalist societies, the belief in values such as the free initiative promotes high degrees of individuation, or high mean values for $f_{P}(x)$. In moral sense, this may be translated into individualism, which is more evident in capitalist societies. However, there is sense of belonging in capitalist societies as well, 
but in certain cases it is an atomized sense. The sense of belonging to parties, religions, groups, and teams 
increases the degree of belonging in capitalist societies. But parties, religions, groups, and teams are not stronger than the modern state in order to promote the sense of belonging. In modernity, state is the strongest way to join or unify people. Nietzsche in \textit{Thus spoke Zarathustra} wrote about the \enquote{new idol}, the modern state:\footnote{\citet[p. 34]{Zarathustra}. Today it is not sufficient to belong to a specific culture, the demand is for states. Terrorists and peoples, such as Catalans, aim to found states.}
\begin{quote}
Somewhere still there are peoples and herds, but not where we live, my brothers: here there are states.

State? What is that? Well then, lend me your ears now, for I shall say my words about the death of peoples.

State is the name of the coldest of all cold monsters. It even lies coldly, and this lie crawls out of its mouth: \enquote{I, the state, am the people.}
\end{quote}
Therefore, according to my interpretation, the modern state generates the highest sense of belonging today. And, following Nietzsche, for societies where state is strong, like the socialist state, the mean value for the process of individuation is low because the modern state \enquote{aspires to the annihilation of the individual.} It is worth
 noting that the modern notions of patriotism and nationalism were 
created in order to increase the sense of belonging even in capitalist societies.\footnote{See \cite{Primoratz}
 for a review on the notions of patriotism and their differences in relation to the notion of nationalism.} But in
  socialist societies the economic system is another key ingredient that amplifies the sense of belonging. In the end,
   that is the reason why the average degree of individuation is
    lower in socialist countries  and the sense of belonging is stronger than in capitalist countries.\footnote{According to \cite{Canovan}, the notion of patriotism may not be sufficient to join people
     in capitalist societies, thus questions about identity and values are also important for the sense of belonging.} 

In human societies, high mean values for the process of individuation ($\bar{f}_{P}(H)\rightarrow 1$) means \enquote{isolated}\footnote{The quotes indicate that a completely isolated individual does not exist. The individual has cultural and collective origins as well.} individuals who do not believe in a cohesive or collective society. Exaggerated individualistic behaviors may lead to situations where \enquote{all against all} is the rule. In general, it is worth noting that high mean values of $f_P(x)$ are not problem in itself. We must consider the associated variance $\sigma_{P}^2$ of the mean psychosocial process of individuation as well. The variance $\sigma_{P}^2$ gives the level of dispersion of the process, offers a measure of \enquote{uniformity} of a specific society or group. The variance is given by
\begin{equation}
\sigma_{P}^2=\frac{1}{N}\sum_{i=1}^{N}\left(f_P(x_i)-\bar{f}_P(S) \right)^2.
\end{equation}
In a hypothetical \enquote{divine society,} for example, we have the maximum mean value for the process of individuation. For such a fictional society, whose members are gods, $f_{P}(x)$ assumes 1 for all individuals. Thus, $P$ plays the role of the universal fuzzy set $U$. Moreover, the variance of the process for this divine society is low, indeed it is zero.\footnote{The Greek Olympus would be a good example for this type of society, but the Mount Olympus has a king, Zeus, then it has hierarchy.} Above all, the variance of the mean value for the process of individuation is a critical problem  in \textit{real} human societies. Such as in the social classes analysis, the problem may be large values for the variance $\sigma_{P}^2$, which is translated into large differences among social classes or high levels of hierarchy. 

It is worth emphasizing that the classification presented here is continuum. Between indicated types in Table 1, which are limits, there are degrees of both individuation and hierarchy. That is the reason why the classification of states, social systems, and societies is not always a two-valued problem, a binary choice between \enquote{yes} or \enquote{no}. China represents such a difficulty today. The difficulty regarding the question whether China is capitalist, or socialist, or communist, or something else, appears due to fuzziness. China possesses degrees of membership in the capitalism and socialism systems. In this example and others, there is no response in terms of an exclusive \enquote{left} or \enquote{right}. Then, fuzzy sets are appropriate tools in order to ban dichotomous points of view in social and political sciences.   

\subsection{Low mean values for the process of individuation}
    
Assuming $f_{P}(x)$ a random variable, we can evaluate its associated variance $\sigma_{P}^2$. Then, societies with low mean values of individuation $\bar{f_{P}}(H)$ are characterized by different values for $\sigma_{P}^2$. For low values of $\sigma_{P}^2$, one has horizontal societies, where the utopian communism is its limit when $\sigma_{P}^2 \rightarrow 0$. On the other hand, a large value of $\sigma_{P}^2$ leads to a vertical society, where socialist dictatorships and real monarchies are examples. As pointed out earlier, $\bar{f}_{P}(H)$ is low for socialist societies (compared to capitalist ones). With high values of  $\sigma_{P}^2$ for this corresponding case, the social organization contains well-defined hierarchies, leading to vertical societies. But, on the other hand, the utopian communism rejects classes and a hierarchical society. That is the reason why $\sigma_{P}^2$ tends to zero for such an ideal society.      
 
 \subsection{High mean values for the process of individuation}

Large values for $\bar{f}_{P}(H)$ indicate other scenarios. For high $\sigma_{P}^2$ for the corresponding choice of $\bar{f}_{P}(H)$, one has vertical societies with supposedly \enquote{isolated} agents. Western capitalist societies are examples for those values of $\sigma_{P}^2$. In this case, even with the patriotic feeling, the sense of belonging to a larger structure is not too emphasized (compared to socialist societies), that is to say, members of this type of social organization give value to individuality to the detriment of collectivity. Such societies present still well-defined hierarchical institutions and differences among classes.  The limit of this case is interpreted as a non-society, i.e., extreme individuality alongside high values for $\sigma_{P}^2$ leads to hordes of Hobbesian \enquote{natural men,} where there is no sense of belonging, and each individual disputes against each individual. In this case, hierarchies are constructed by differences in power, or by using a Nietzschean concept, by differences among wills to power. 

The last limit to be studied is the case where  $\bar{f}_{P}(H)$ is high and $\sigma_{P}^2$ is low. In this curious case, the limit of $\sigma_{P}^2$ for such a society promotes an organization of gods, i.e., each member is view as a complete individual (a fictitious individual), and there is lack of both classes and well-defined hierarchical organizations (see Table 1 for an overview of types of societies).

\begin{table}
\begin{center}
\begin{tabular}{| c | c |}
\hline
\textbf{Mean Value of Individuation} $f_{P}(x)$ & \textbf{Variance}: $\sigma_{P}^2$  \\ \hline
                                               & $\sigma_{P_{high}}^2$ $\rightarrow$ \enquote{all against all} (non-society) \\ 
$\bar{f}_{P_{high}}(H)$     &                                                                                                  \\
                                               & $\sigma_{P_{low}}^2$ $\rightarrow$ divine society \\ \hline
                                               
                                               & $\sigma_{P_{high}}^2$ $\rightarrow$ vertical society (hierarchy) \\ 
$\bar{f}_{P_{low}}(H)$     &                                                                                                   \\
                                               & $\sigma_{P_{low}}^2$ $\rightarrow$ horizontal society (non-hierarchy) \\
                                               
\hline

\end{tabular}
\caption{Types of human $(H)$ societies according to the values of the psychosocial process of individuation and their corresponding variances. The table illustrates possible configurations and their limits. There are degrees between the illustrated limits.}
\end{center}
\end{table}

\section{Final remarks}
Fuzzy sets are one of the most revolutionary concepts in science and technology. In this paper, I adopted this idea to look
 at the problem of individuation. Considered as a two-valued problem, the process of individuation has been thought of
  as a process that offers only two answers: non-individual and individual. Rather than a process described by an ordinary
   set, I proposed a process of individuation described by a fuzzy set. Then, its corresponding membership function
    assumes a continuum interval $[0,1]$ as image. With this interpretation, the process of individuation in its
     psychosocial aspect presents degrees of individuation: the closer to 0, the smaller the degree of individuation, and
      the closer to 1, the greater the degree of individuation. Each species has their specific values of individuation. Due
       to the possible different values of variance for the individuation in human societies, the psychosocial process of
        individuation may indicate both degrees of cohesion and hierarchy. For high values of variance, for example,
         the psychosocial process of individuation describes both dictatorial societies and hierarchical societies. 
         For low values of variance, one has horizontal societies without (or almost without) both hierarchical
          structures and classes.    

The psychosocial process of individuation proposed in this article may offer new variables in order to describe human societies and social groups, with fuzzy sets indicating degrees of both individuation and belonging.

\section*{Acknowledgments}
This study was financed in part by the Coordenação de Aperfeiçoamento de Pessoal de Nível Superior---Brasil (CAPES)---Finance Code 001. I would like to thank Alexandre Patriota for helpful suggestions, Elena Senik for reading the manuscript and an anonymous referee for important questions raised during the review process.

\end{document}